\documentclass{ws-ijmpa}
\usepackage[super,compress]{cite}
\usepackage{graphicx}
\usepackage{cancel}
\usepackage{cite}
\usepackage{slashed}
\usepackage{amsmath,bbm}
\usepackage{amssymb}
\usepackage{graphicx}
\usepackage[latin1]{inputenc}
\usepackage{mathrsfs}
\usepackage{hyperref}

\newcommand{\bea}{\begin{eqnarray}}
\newcommand{\eea}{\end{eqnarray}}
\newcommand{\be}{\begin{equation}}
\newcommand{\ee}{\end{equation}}
\newcommand{\ba}{\begin{array}}
\newcommand{\ea}{\end{array}}

\def\gsim{\mathrel{\rlap{\lower4pt\hbox{\hskip1pt$\sim$}}
    \raise1pt\hbox{$>$}}}

\begin{document}
\markboth{Matthew Reece}{Physics at a Higgs factory}

%
\catchline{}{}{}{}{}
%

\title{Physics at a Higgs Factory}

\author{Matthew Reece\footnote{E-mail: mreece@physics.harvard.edu} }

\address{Department of Physics, Harvard University,\\
17 Oxford St., Cambridge, MA 02138 USA}

\maketitle


\begin{abstract}
I give an overview of the physics potential at possible future $e^+ e^-$ colliders, including the ILC, FCC-ee, and CEPC. The goal is to explain some of the measurements that can be done in the context of electroweak precision tests and Higgs couplings, to compare some of the options under consideration, and to put the measurements in context by summarizing their implications for some new physics scenarios. This is a writeup of a plenary talk at the Hong Kong University of Science and Technology Jockey Club Institute for Advanced Study Program on High Energy Physics Conference, January 18--21, 2016. Some previously unpublished electroweak precision results for FCC-ee and CEPC are included.

\end{abstract}



\section{Introduction}	

One of the most exciting developments in high-energy physics in recent years is the design and planning of multiple large-scale future experiments. These include both electron-positron and proton-proton colliders. A major goal of the electron-positron machines is to precisely measure the couplings of the Higgs boson. For this reason they are often referred to as ``Higgs factories,'' and all of the machines being planned will run at energies near 240 GeV where the $e^+ e^- \to Z h$ cross section is largest. This note grew out of a talk that I was asked to give at the Hong Kong IAS on the topic ``Physics at a Higgs Factory.'' One of the major physics questions, especially for the CEPC collider whose planning is still at an early stage, is the extent to which the collider should be solely focused on the Higgs. How important is $Z$-pole physics? How important are measurements on the $t{\bar t}$ or $W^+ W^-$ threshold? In this note I will explain some of the physics that I think is most useful for making informed decisions about such questions. I also want to give some context. Rather than just asking about how accurately measurements can be done, I will ask: what will these measurements tell us about what could lie beyond the Standard Model? The talk was based in part on previous studies in collaboration with JiJi Fan and Lian-Tao Wang on electroweak precision observables at future colliders\cite{Fan:2014vta,Fan:2014axa,CEPC-SPPCStudyGroup:2015csa}, as well as work by others that I will cite below. I have also added some new material in light of discussions at the Hong Kong meeting and other recent workshops.

\section{Electroweak Precision}

\subsection{Projected Reach in S and T}

I will focus my discussion on a few dimension-six operators that involve the Higgs boson and generally receive important contributions in natural theories. For instance, in a supersymmetric theory, the stops that cancel the leading quadratic divergence also run in the loop to produce these operators. Namely,
\begin{align}
S~{\rm parameter}: &\quad S \left(\frac{\alpha}{4s_W c_W v^2}\right) h^\dagger \sigma^i h W^i_{\mu \nu}B^{\mu \nu} \\
T~{\rm parameter}: &\quad -T \left(\frac{2\alpha}{v^2}\right) \left|h^\dagger D_\mu h \right|^2 \\
{\rm Higgs~decays}: &\quad c_{hgg} h^\dagger h G^a_{\mu \nu}G^{a\mu\nu} + c_{h\gamma\gamma} h^\dagger h F_{\mu \nu}F^{\mu \nu}. \label{eq:higgsdecay}
\end{align}
Of course, there are many other dimension six operators and it is interesting to constrain them all.\cite{Ellis:2015sca,Ge:2016zro,deBlas:2016ojx} Flavor-violating operators, for example, can be leading probes of new physics. But this set of operators is very common in any new physics coupling to the Higgs boson. Other familiar operators in the electroweak sector tend to be subdominant: the TGC operator $W^3$ and the $W$-parameter $(DW)^2$, for instance, have tiny coefficients when SU(2) multiplets are integrated out at one loop, while the $U$-parameter is dimension 8.

\begin{figure}
\begin{center}
\includegraphics[width=0.5\textwidth]{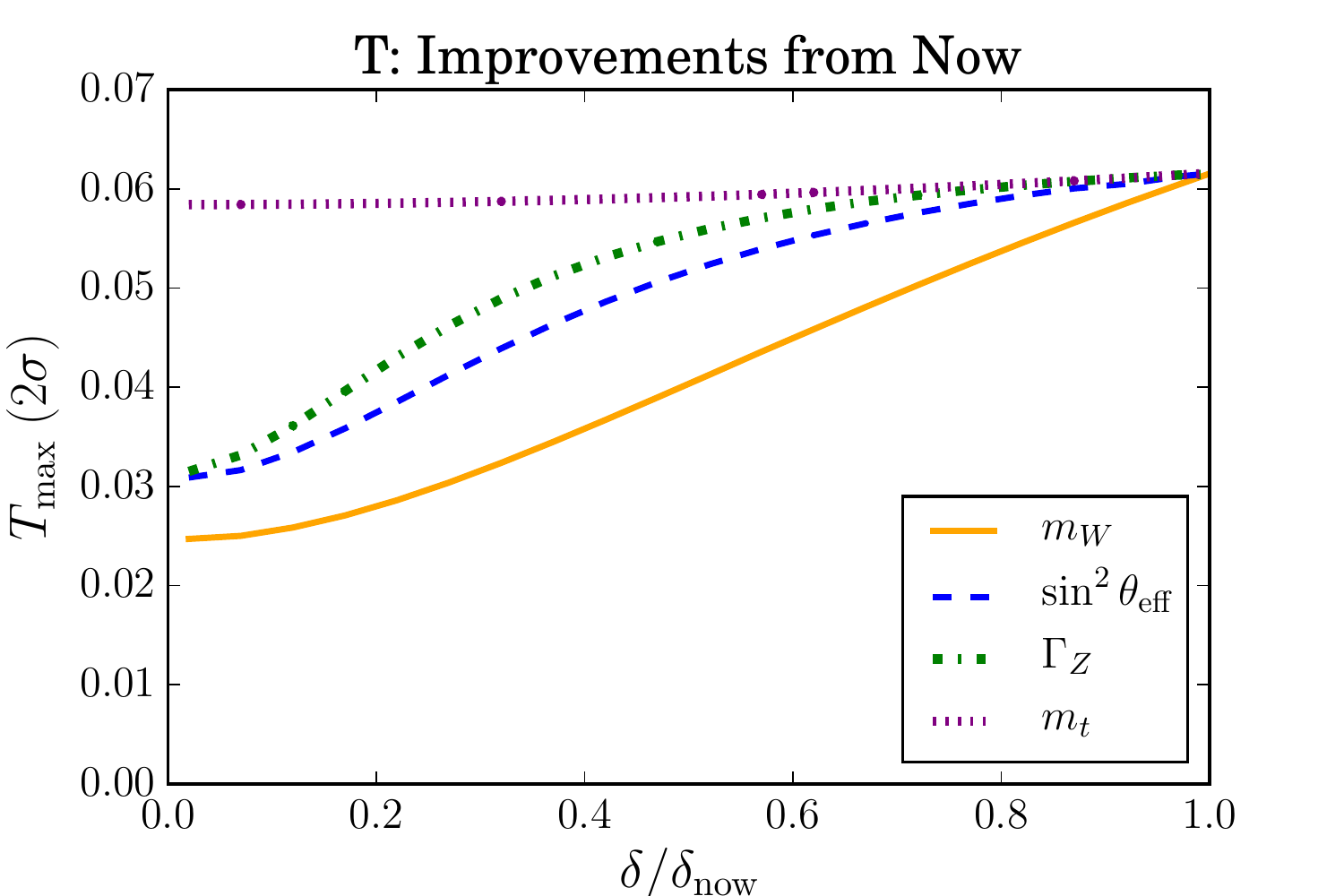}\includegraphics[width=0.5\textwidth]{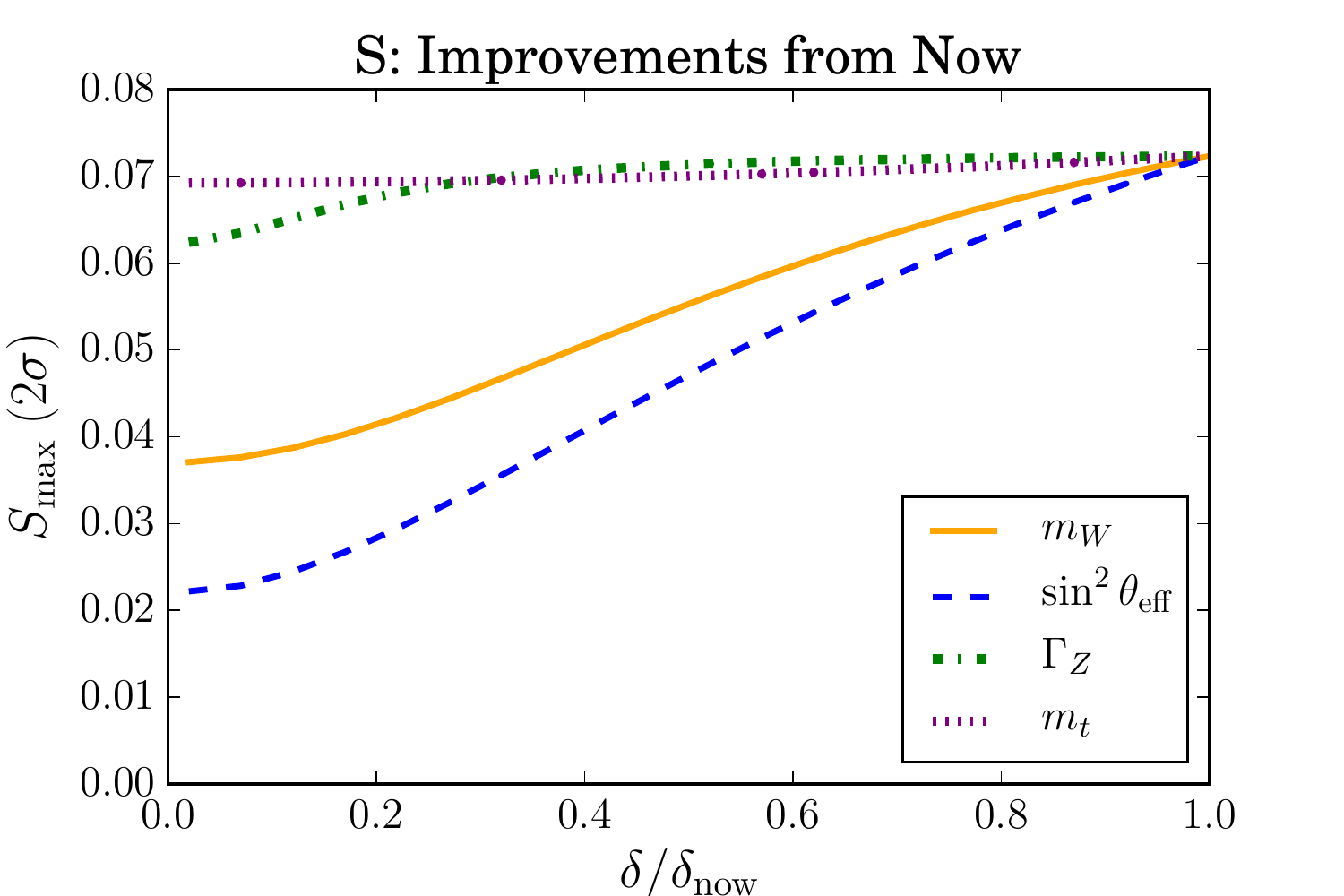}
\end{center}
\caption{Effects on the current $T$ and $S$ parameter constraints of reducing uncertainties on individual quantities $m_W$, $\sin^2 \theta_{\rm eff}$, $\Gamma_Z$, and $m_t$. In this plot $\delta$ includes both experimental and theoretical uncertainties. In the $T$ plot, we have set $S = 0$ (performing a one-parameter fit), and vice versa.}
\label{fig:reduceuncertaintiescurrent}
\end{figure}

The electroweak precision fit depends on a number of experimental inputs. Among the most important ones to improve are the $W$ mass, the effective weak mixing angle $\sin^2 \theta_{\rm eff}$ (measured through quantities like left-right asymmetries), the top mass, and the $Z$ mass and width. A first question we can ask is: given our {\em current} knowledge of these observables, which ones are the bottlenecks? In other words, which quantities are the most important ones to improve our knowledge of {\em first} if we want a better electroweak fit? The plots in Figure \ref{fig:reduceuncertaintiescurrent} answer this question for the $T$ and $S$ parameters. The most effective way to improve the bound on the $T$ parameter is to obtain a better measurement of the $W$ boson mass. This can be done in Higgs factories operating at 240 GeV, where $W$ pairs can be produced in abundance. It does not require $Z$ pole physics. On the other hand, improving the bound on the $S$ parameter demands better measurements of $\sin^2 \theta_{\rm eff}$. For this, high luminosity on and around the $Z$ pole is the preferred strategy. This is one motivation for operating future $e^+ e^-$ colliders near the $Z$ pole: repeating much of the LEP physics program with higher precision can significantly improve our knowledge of the $S$ parameter. (That said, I don't know of studies investigating how well we could extract $\sin^2 \theta_{\rm eff}$ from observables at 240 GeV.) Figure \ref{fig:reduceuncertaintiescurrent} also illustrates that improvements saturate at some point. If we measure the $W$ boson mass a factor of 5 or so better than we now know it---that is, to a precision of about 3 MeV, which CEPC, for instance, would accomplish---it is no longer the bottleneck in our knowledge of the $T$ parameter, and we might then want to improve measurements of other quantities like $\Gamma_Z$ and $m_t$ as secondary priorities.

The leading observables that matter for probing $S$ and $T$ will be measured with significantly higher precision than we have now at {\em any} of the $e^+ e^-$ colliders under discussion. The ILC will measure\cite{Baak:2014ora} the $W$ mass to 5 MeV and $\sin^2\theta_{\rm eff}$ to $1.3 \times 10^{-5}$; CEPC will measure\cite{CEPC-SPPCStudyGroup:2015csa} the $W$ mass to 3 MeV and $\sin^2\theta_{\rm eff}$ to $2.3 \times 10^{-5}$; and the FCC-ee will (according to more conservative estimates\cite{Baak:2013fwa}) measure the $W$ mass to 1.2 MeV and $\sin^2\theta_{\rm eff}$ to $0.3 \times 10^{-5}$. When the experimental uncertainties become particularly small, theory uncertainties matter a great deal as well, and we have included estimates of the remaining theory uncertainty after 3-loop calculations are performed \cite{Freitas:2000gg,Awramik:2002wn,Awramik:2006uz,Freitas:2014hra}. (Such calculations will be a crucial task for theorists to complete in the coming years.)

\begin{figure}
\begin{center}
\includegraphics[width=\textwidth]{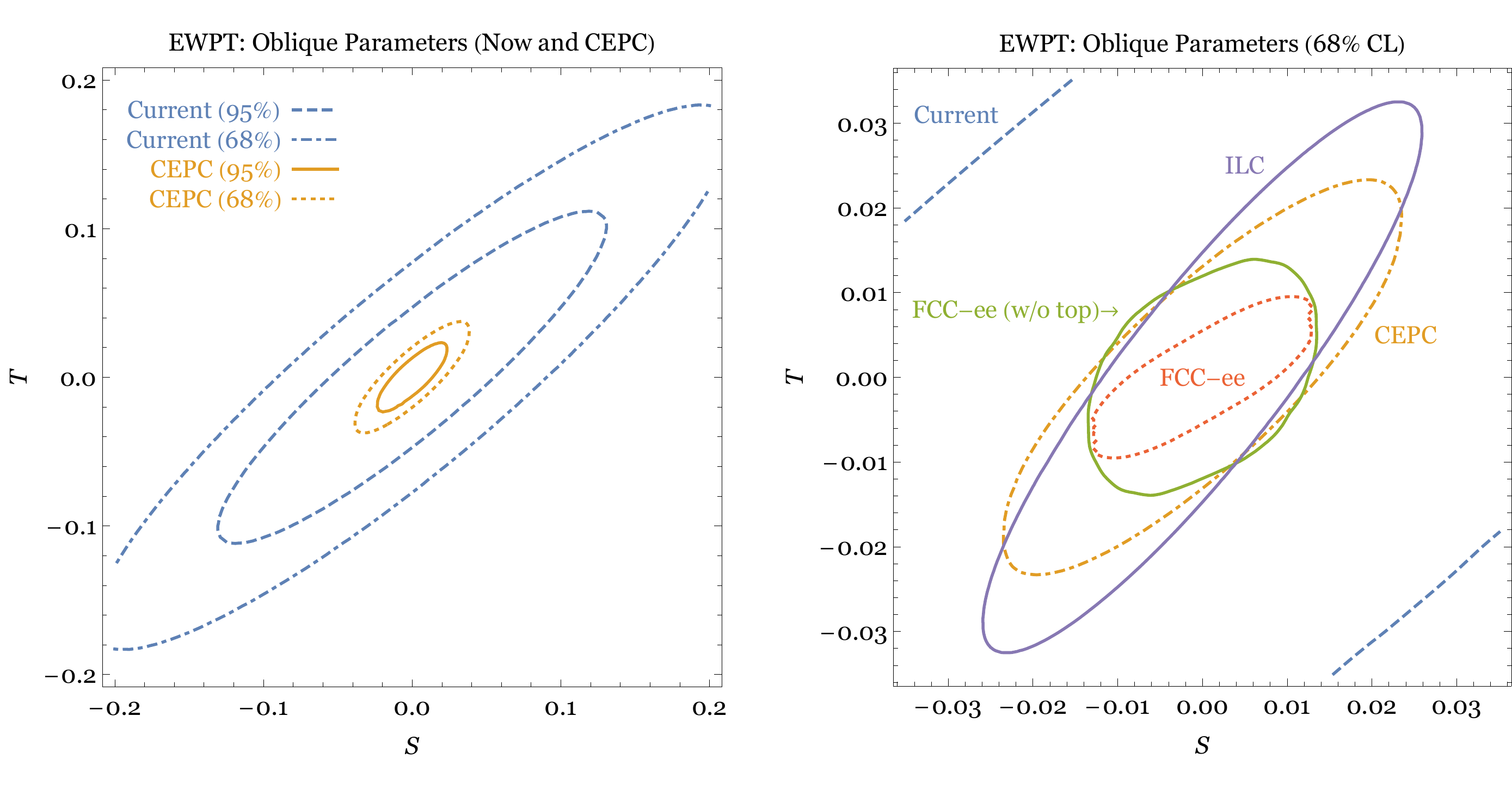}
\end{center}
\caption{Precision that will be achieved for the $S$ and $T$ parameters at future colliders. At left: comparison of the current electroweak precision fit (artificially recentered at $S = T = 0$) with the expectations for CEPC, using projections from the pre-CDR.\cite{CEPC-SPPCStudyGroup:2015csa} At right: 68\% contours for current data, CEPC, ILC, FCC-ee without running at the top threshold, and FCC-ee with running at the top threshold. (The latter two fits were referred to as ``TLEP-W'' and ``TLEP-t'' in our previous work.\cite{Fan:2014vta})}
\label{fig:STfits}
\end{figure}

The projected fits in the $(S,T)$ plane for the various future experiments are shown in Fig.~\ref{fig:STfits}. The ILC and CEPC are projected to do comparably well, although they have slightly different strengths. FCC-ee is more ambitious in terms of its projected mass resolution and luminosity, with correspondingly better projected fits.

\subsection{Experimental Choices}

\begin{figure}
\begin{center}
\includegraphics[width=0.5\textwidth]{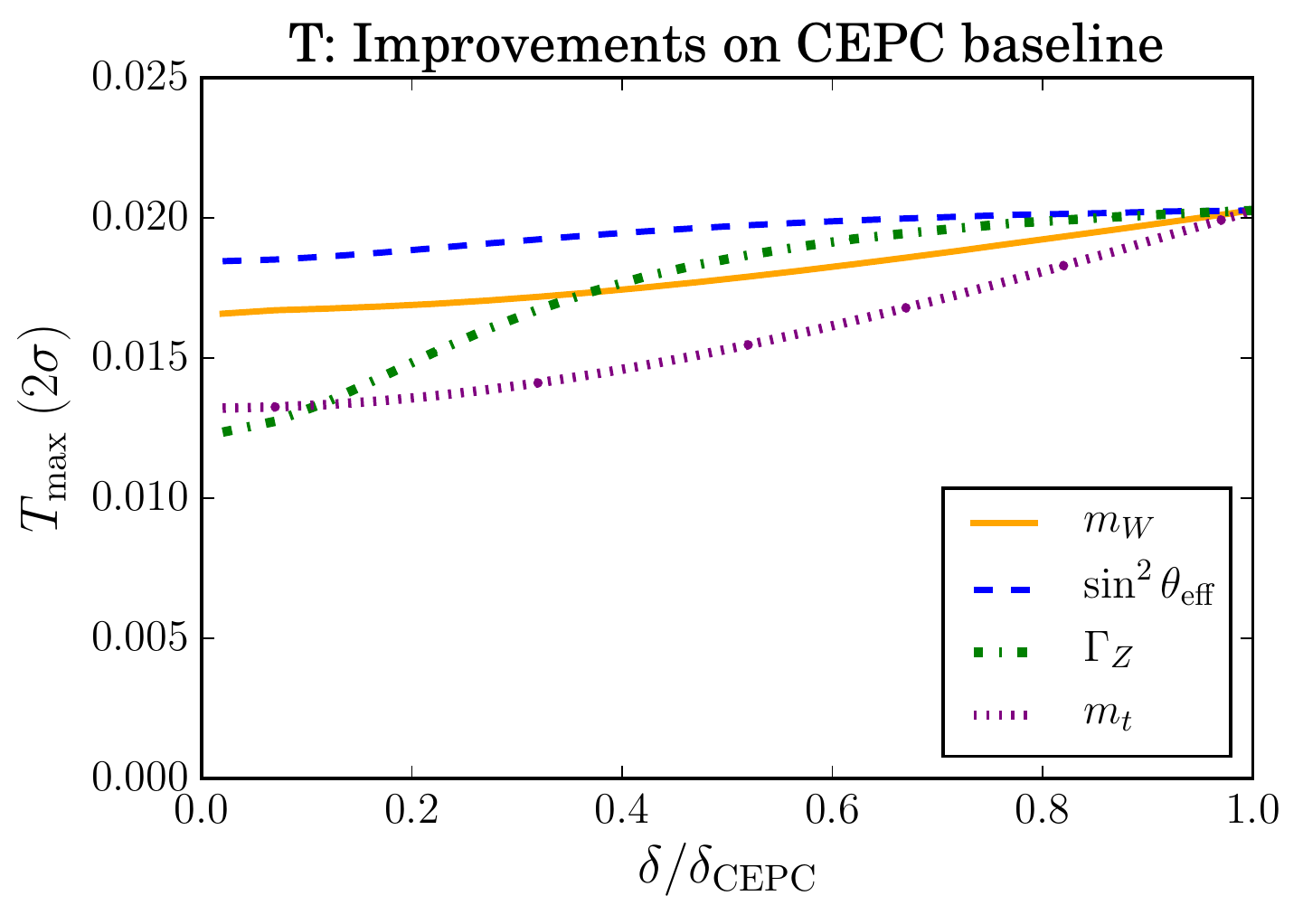}\includegraphics[width=0.5\textwidth]{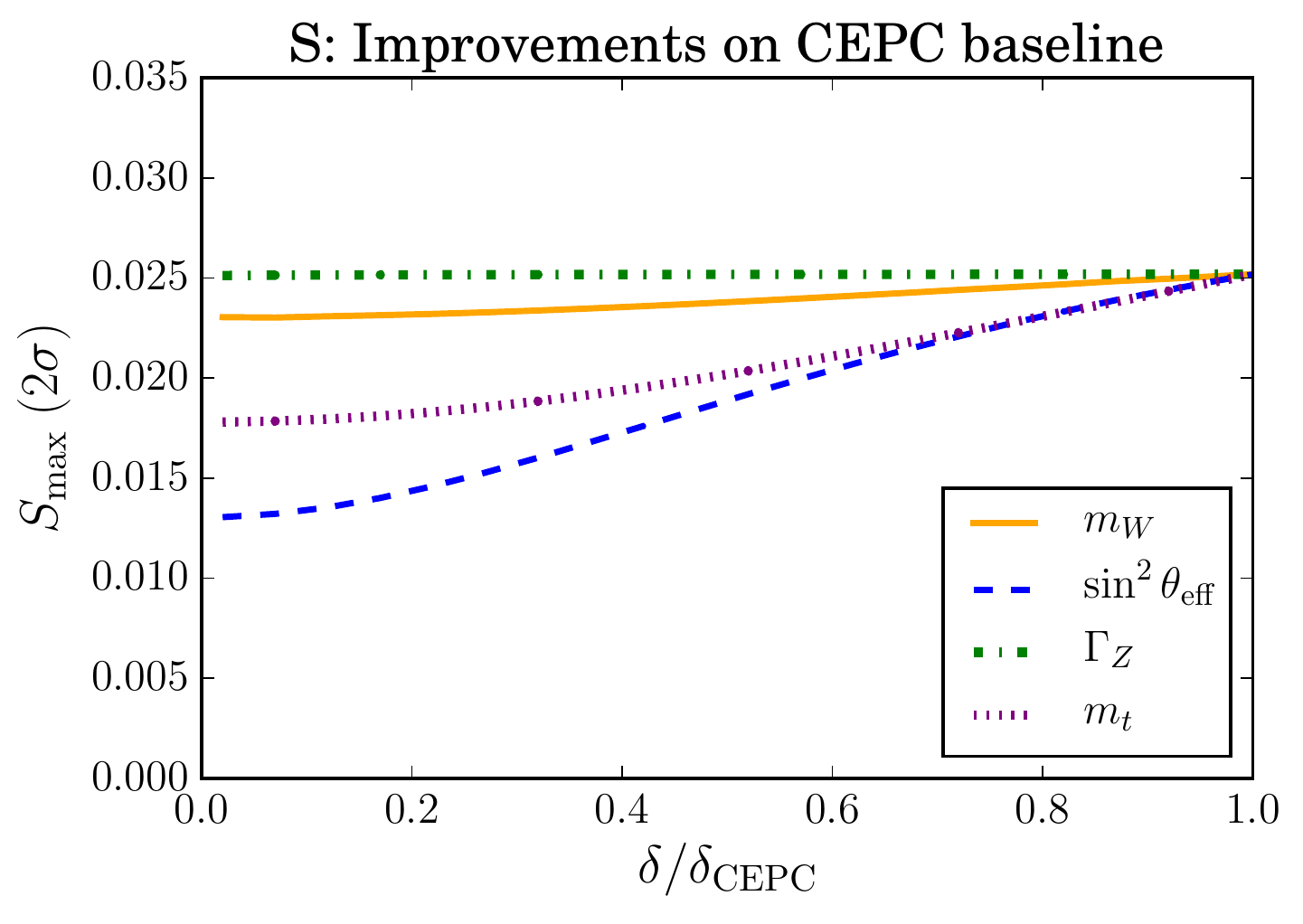}
\end{center}
\caption{Like Figure \ref{fig:reduceuncertaintiescurrent}, except relative to the baseline CEPC measurements (as tabulated in Table 4.5 of the pre-CDR\cite{CEPC-SPPCStudyGroup:2015csa}) rather than to current measurements.}
\label{fig:reduceuncertaintiesCEPC}
\end{figure}

Now that we have seen the estimates of how well the different colliders can do, it's useful to step back and ask which features of the colliders lead to these results, and which observables are the most important ones to optimize. We have seen that the first step is to produce more precise measurements of $m_W$ and $\sin^2 \theta_{\rm eff}$. What next? To help answer this question, in Figure \ref{fig:reduceuncertaintiesCEPC} we have plotted the change in the $S$ and $T$ parameter constraints if we begin with the CEPC baseline design and then improve one measurement at a time. We see that once the CEPC baseline is achieved---which includes a measurement of $m_W$ to 3 MeV accuracy---the most efficient way to improve the $T$ parameter fit is to obtain a better measurement of the top quark mass. To improve the $S$ parameter fit, we should either improve $\sin^2 \theta_{\rm eff}$ (which, as discussed in the CEPC pre-CDR, is likely possible for at least a factor of 2 beyond the baseline design) or the top quark mass. Further improvements in $m_W$ are of limited use and $\Gamma_Z$ can be useful only with dramatic improvements (at which point, because it depends on a different linear combination of $S$ and $T$ than other observables, it can yield a better bound).

The CEPC collider design does not, at this time, include a plan to run at the top threshold. FCC-ee does, with corresponding improvements in precision for the $T$ parameter. Top mass measurements at $e^+ e^-$ colliders can be much more precise than at the LHC, because the threshold scan can determine the top mass and width in the 1S scheme which is less subject to theoretical uncertainties than kinematic measurements of the mass.\cite{Hoang:2013uda} One argument in favor of a linear collider is that it is easier to go to the top threshold and do precision measurements there---and to go to even higher energies and measure couplings of the top quark to heavy gauge bosons and the Higgs boson.

On the other hand, refining the $Z$ mass and width measurements can also be a route to higher accuracy in the electroweak fits once the initial bottlenecks are over come. Here the circular colliders have an advantage: they have a useful energy calibration based on resonant spin depolarization.\cite{ALEPH:2005ab} The idea is that polarized beams will precess in the magnetic field, so applying an orthogonal field at the right frequency can depolarize the electrons. Similar to NMR, this technique allows a very precise energy calibration.

\begin{table}[ph]
\tbl{FCC-ee measurement uncertainties. These are drawn from a number of references.\cite{Gomez-Ceballos:2013zzn,Lepage:2014fla,Baak:2013fwa,Baak:2014ora,Awramik:2003rn,Awramik:2006uz,Freitas:2014hra} Further explanation of the ``TLEP-$t$'' column, which is largely based on the Snowmass Electroweak Working Group report and used for the inner FCC-ee contour in Figure \ref{fig:STfits}, may be found in our earlier work.\cite{Fan:2014vta} The ``stat.'' and ``syst.'' columns draw more heavily from the TLEP ``First Look'' report,\cite{Gomez-Ceballos:2013zzn} and are used in Figure \ref{fig:fcceefits}.}
{\begin{tabular}{|c|c|c|c|c|}
\hline
& ``TLEP-$t$'' & Stat. & Stat.+Syst. & Theory \\
\hline
$\alpha_s(M_Z^2)$ &  $\pm 1.0 \times 10^{-4}$ &  $\pm 1.0 \times 10^{-4}$ &  --- &  --- \\
$\Delta\alpha_{\rm had}^{(5)}(M_Z^2) $ & $\pm 4.7 \times 10^{-5}$ &  $\pm 4.7 \times 10^{-5}$ & --- & --- \\
$m_Z$ & $\pm 100$~keV & $\pm 5$~keV & $\pm 100$~keV & --- \\
$m_t$ & $(\pm 0.02_{\rm exp} \pm 0.1_{\rm th})$ GeV & $\pm 10~{\rm MeV}$ & $\pm 14~{\rm MeV}$ & $\pm 100~{\rm MeV}$ \\
$m_h$ & $\pm 0.1~{\rm GeV}$ & $\pm 0.1~{\rm GeV}$ & --- & --- \\
\hline
$m_W$ & $\left(\pm 1.2_{\rm exp} \pm 1_{\rm th}\right)~{\rm MeV}$ & $\pm 300~{\rm keV}$ & $\pm 500~{\rm keV}$ & $\pm 1~{\rm MeV}$ \\
$\sin^2\theta^{\ell}_{\rm eff}$  & $\left(\pm 0.3_{\rm exp} \pm 1.5_{\rm th}\right) \times 10^{-5}$ & $\pm 1.0 \times 10^{-6}$ & $\pm 1.0 \times 10^{-6}$ & $\pm 1.5 \times 10^{-5}$ \\
$\Gamma_Z$ & $\left(\pm 1_{\rm exp} \pm 0.8_{\rm th}\right) \times 10^{-4}~{\rm GeV}$ & $\pm 8~{\rm keV}$ & $\pm 100~{\rm keV}$ & $\pm 80~{\rm keV}$ \\
\hline
\end{tabular} \label{tabfcc}}
\end{table}

\begin{figure}
\begin{center}
\includegraphics[width=0.6\textwidth]{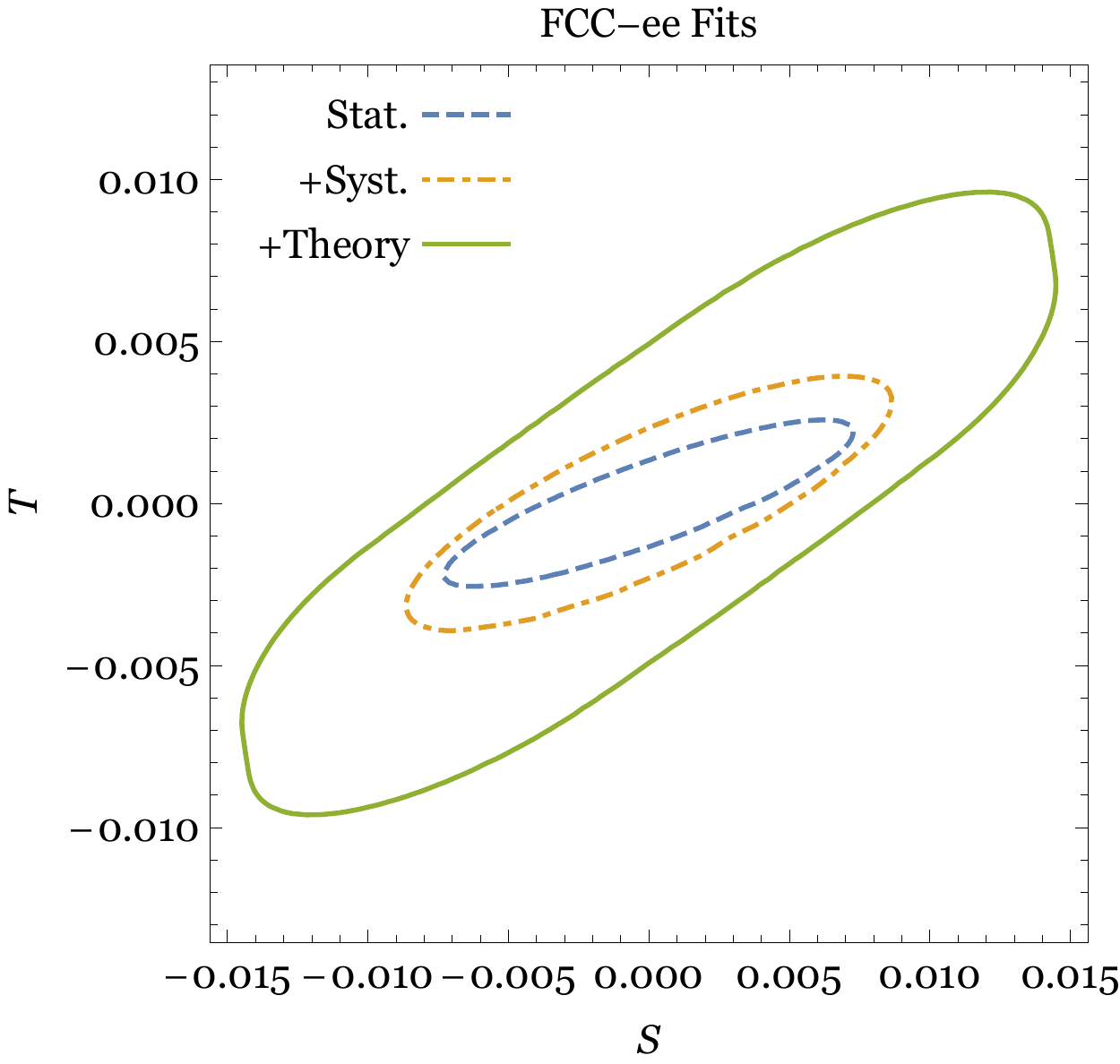}
\end{center}
\caption{68\% CL contours in the $S-T$ plane for various estimates of the capabilities of FCC-ee. The dashed blue and dot-dashed orange contours are based on purely statistical uncertainties and statistical plus systematic uncertainties from the TLEP ``First Look'' report.\cite{Gomez-Ceballos:2013zzn} The green contour also adds in theoretical uncertainties on $m_t, m_W, \sin^2 \theta_{\rm eff},$ and $\Gamma_Z$ as in our earlier work\cite{Fan:2014vta} and is almost indistinguishable from the inner FCC-ee curve shown in Figure \ref{fig:STfits}.}
\label{fig:fcceefits}
\end{figure}

Both CEPC and FCC-ee can take advantage of this precise energy calibration to measure $m_Z$ and $\Gamma_Z$---and possibly other quantities like $m_W$---more accurately than the ILC. The FCC-ee electroweak fits we have presented in the past\cite{Fan:2014vta} largely come from projections in the Snowmass Electroweak Working Group studies\cite{Baak:2013fwa} that are typically more conservative than those in the TLEP ``First Look'' report\cite{Gomez-Ceballos:2013zzn}. For instance, resonant spin depolarization might allow the $W$ mass to be measured to 500 keV accuracy or better. (This could be a major motivation for accumulating luminosity on the $WW$ threshold at $\sqrt{s} \approx 160$ GeV.) On the other hand, theoretical uncertainties might ultimately limit the utility of high experimental precision. To shed some light on this issue, we have plotted in Figure \ref{fig:fcceefits} the $(S,T)$ ellipses that would be achieved with purely statistical and systematic uncertainties---which are exquisitely precise, especially for the $T$-parameter---and the effect of also including theoretical systematics. The inputs to these fits are tabulated in Table \ref{tabfcc}. Depending on one's point of view, this is either an indication that aggressive estimates of the experimentally achievable precision are overly optimistic or that theorists will have to work harder to overcome significant obstacles to higher-precision calculations.

The FCC-ee project has continued to investigate the possibility of more high precision measurements. The experimental issues involved in resonant depolarization for energy calibration at FCC-ee were recently studied systematically.\cite{Koratzinos:2015hya} Other continuing studies at FCC-ee include the measurement of top quark couplings\cite{Janot:2015yza} and of the value of $\alpha$ at the $Z$ mass scale,\cite{Janot:2015gjr} avoiding the need to directly understand hadronic contributions to the running. Such studies, beyond the preliminary estimates, are important to assess whether bottlenecks can be overcome to achieve higher precision results like the inner ellipses in Figure \ref{fig:fcceefits}.

\subsection{Summary of the $(S,T)$ Fits}

Of course, we want to have the best measurements possible of many different quantities. But as a reasonable set of baselines that we should ask for from future experiments, we suggest:
\begin{itemize}
\item Measure $m_W$ to better than 5 MeV. The current uncertainty is 15 MeV. All designs being discussed meet this standard.
\item Measure $\sin^2 \theta_W$ to better than $2 \times 10^{-5}$. The current uncertainty is $16 \times 10^{-5}$. Again, all designs being discussed can deliver this.
\item Measure $m_Z$ and $\Gamma_Z$ to 500 keV precision (currently 2 MeV). The future circular colliders would deliver this accuracy, but the ILC would not.
\item Measure $m_t$ to 100 MeV precision (currently somewhere around 0.8 GeV, with difficult-to-quantify theoretical uncertainties). The ILC and FCC-ee promise this accuracy, but CEPC does not.
\item Have precise enough theory to make use of these results. This requires at least 3-loop calculations.
\end{itemize}
Each of the possible future electron-positron colliders delivers at least a substantial subset of this wish list (and some do far better for some quantities!), and would provide order-of-magnitude improvements on our current knowledge of electroweak precision. The circular and linear colliders would have interesting complementarity if both are constructed.

\begin{figure}
\begin{center}
\includegraphics[width=0.6\textwidth]{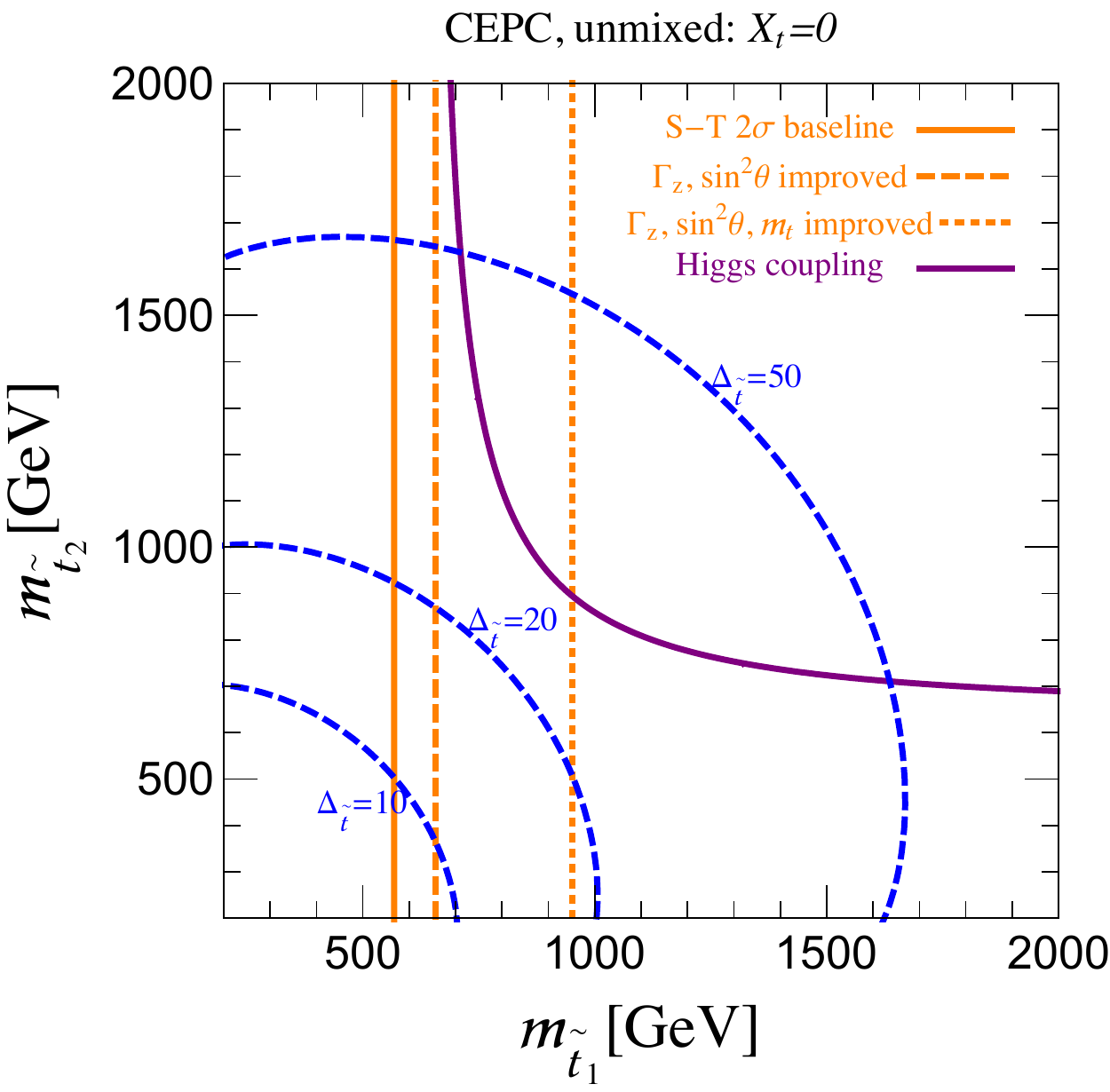}
\end{center}
\caption{Probing stop masses with measurements at CEPC. The orange curves are the reach for the CEPC baseline electroweak precision fit and the extensions if $\Gamma_Z$ and $m_t$ measurements are improved as well. The purple curve comes from measurements of Higgs couplings to photons and gluons. The blue dashed lines indicate fine-tuning levels.}
\label{fig:susycepc}
\end{figure}

As one illustration of how to put the $S$ and $T$ parameter numbers in context, we show in Figure \ref{fig:susycepc} the translation of these bounds into constraints on stop soft masses. The electroweak fit---mostly the $T$-parameter---translates into a constraint on the left-handed stop soft mass of around 700 GeV. This is comparable to already existing direct search bounds from the LHC, but more robust. Direct search bounds can be evaded if the neutralinos are heavy or if the stop decays in unusual ways. Future electron-positron colliders are nicely complementary, probing particles like the stop directly through their coupling to the Higgs boson and definitively closing loopholes. The blue dashed lines in Figure \ref{fig:susycepc} show that these measurements would constrain supersymmetry to be tuned down to at least the few percent level. There is an interesting ``blind spot'' for stops with a left-right mixing $X_t \approx \sqrt{m_{{\tilde t}_{\rm heavy}}^2 - m_{{\tilde t}_{\rm light}}^2}$, for which the lighter stop mass eigenstate decouples from the Higgs boson, but in this limit it also plays no role for naturalness so the basic conclusion is unchanged.\cite{Fan:2014axa,Craig:2014una} Furthermore, other precision observables like $b \to s\gamma$ limit the most natural parts of the blind spot parameter space.

\section{Higgs Boson Measurements}

\subsection{The Central Higgs Factory Physics: Coupling Measurements}

Of course, a major part of the precision physics program at a Higgs factory is measuring the Higgs couplings! This offers exciting opportunities to probe new physics, in addition to testing a portion of the Standard Model that even after the high-luminosity LHC we will still have only coarse information about. Unlike the case of electroweak precision, {\em all} of the $e^+ e^-$ colliders under consideration will pursue a very similar physics program at 240 GeV (so there are fewer contrasts to draw between different choices). For that reason I will focus less on comparing the options and more on explaining how they constrain various new physics scenarios.

\begin{figure}
\begin{center}
\includegraphics[width=0.7\textwidth]{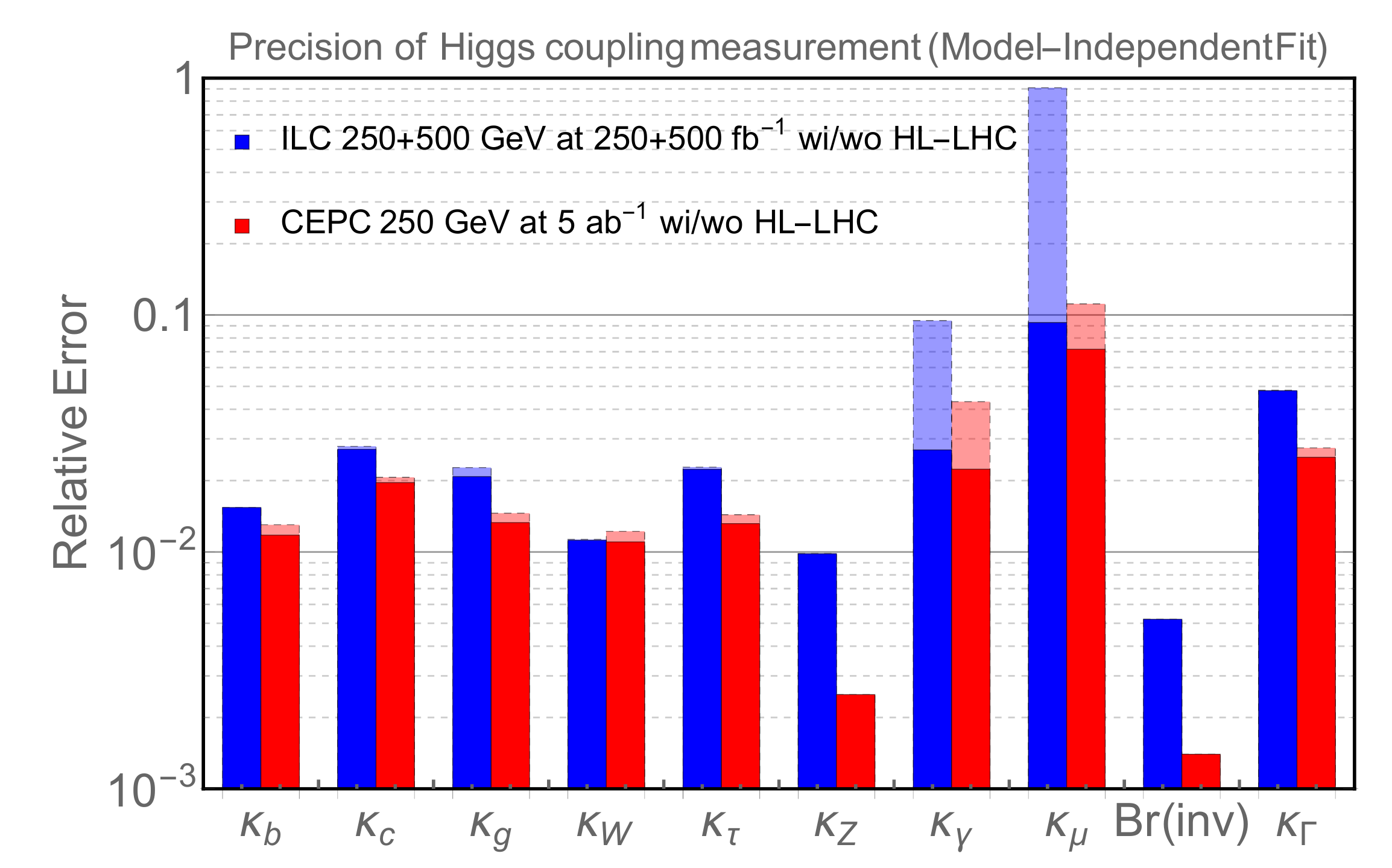}
\end{center}
\caption{Uncertainties on Higgs couplings from a 10-parameter fit at CEPC and ILC, lifted directly from Figure 3.20 of the CEPC pre-CDR \cite{CEPC-SPPCStudyGroup:2015csa}. The bars with dashed edges include a combination with the high-luminosity LHC.}
\label{fig:higgscouplings}
\end{figure}

The prospects for Higgs coupling measurements at future $e^+ e^-$ colliders have been extensively studied.\cite{Peskin:2012we,Asner:2013psa,Han:2013kya,Dawson:2013bba,Peskin:2013xra,Bechtle:2014ewa,Fan:2014vta,CEPC-SPPCStudyGroup:2015csa,Han:2015ofa,Craig:2015wwr} A summary of the precision that will be attained is shown in Figure \ref{fig:higgscouplings}, which is extracted from the CEPC pre-CDR and offers a comparison of measurements at CEPC and at the ILC. The figure also shows that the high-luminosity LHC can complement a future lepton collider, for instance by obtaining a precise measurement of the ratio $\Gamma(h \to \gamma\gamma)/\Gamma(h \to ZZ^*)$. The summary is that all of the largest Higgs couplings will be measured to roughly percent-level accuracy. Because Higgs factories dominantly rely on the Higgsstrahlung process $e^+ e^- \to Z^* \to Zh$, the coupling to the $Z$ boson will be especially well-measured, particularly at circular colliders with their very large projected luminosity. Even the small coupling to photons will be measured to within a few percent accuracy (folding in knowledge from LHC), and the tiny coupling to muons to within about ten percent. Any of the future $e^+ e^-$ colliders under consideration will take us from the LHC's fuzzy, out-of-focus picture of a ``Standard Model-like'' Higgs to a truly precise knowledge that the Higgs is (or, more excitingly, is not!) as predicted.

\begin{figure}
\begin{center}
\includegraphics[width=0.6\textwidth]{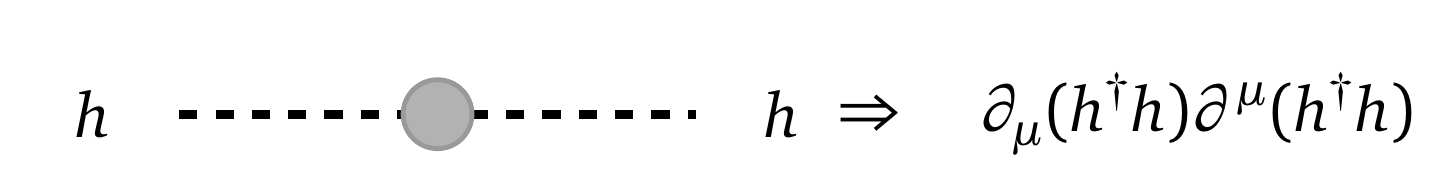}
\end{center}
\caption{Any new physics coupling to the Higgs boson can enter in a loop---here represented by an abstract blob---and produce a wavefunction renormalization effect through the dimension-six operator shown. This affects the rate for any Higgs-related process, including the Higgsstrahlung process that can be exquisitely measured.\cite{Englert:2013tya,Craig:2013xia}}
\label{fig:wavefunction}
\end{figure}

The coupling of the $Z$ to the Higgs will be very precisely measured through the Higgsstrahlung cross section itself. This single measurement is already a powerful probe of new physics \cite{Englert:2013tya,Craig:2013xia,Henning:2014gca,Craig:2014una}. As shown in Figure \ref{fig:wavefunction}, any new physics coupling to the Higgs will lead to a wavefunction renormalization effect that will affect the overall rate of Higgs production. In theories like Twin Higgs models\cite{Chacko:2005pe}, Folded Supersymmetry\cite{Burdman:2006tz}, or other realizations of ``neutral naturalness,'' the new physics coupling to the Higgs can be difficult to see directly. It is encouraging that even in the most pessimistic such scenario---new singlet scalars coupling to the Higgs portal---the Higgsstrahlung precision achievable at CEPC or FCC-ee can probe masses up to several hundred GeV and provide a very general test of fine-tuning \cite{Craig:2013xia}. This is a useful complement to more model-specific tests, which are often more precise when they are available. For instance, in the case of Folded Supersymmetry, the folded stops have the same electroweak quantum numbers as normal stops, and the strongest probe of these particular models at a future $e^+ e^-$ collider is likely to be the $T$-parameter \cite{Fan:2014axa}.

Another example of a scenario that can be tightly constrained by the Higgsstrahlung rate is the case of a composite Higgs \cite{Contino:2010rs}. In these models the Higgs is a pseudo-Nambu-Goldstone boson with decay constant $f$, and the potential has a form like
\begin{equation}
V(h) \sim \frac{a\lambda^2}{16\pi^2} \cos(h/f) + \frac{b\lambda^2}{16\pi^2} \sin^2(h/f),
\end{equation}
where $\lambda$ is some spurion for shift-symmetry breaking. An obvious prediction is that $v \sim f$, so all such models are fine-tuned by adjusting $a = 2b(1+\epsilon)$ to produce a small VEV $\left<h\right>^2 \approx 2 \epsilon f^2$. Composite Higgs models predict modified couplings
\begin{equation}
\frac{g_{VVh}}{g^{\rm SM}_{VVh}} = \sqrt{1 - \frac{v^2}{f^2}}.
\end{equation}
Thus, the Higgsstrahlung measurement has a very direct interpretation as a measure of fine-tuning. FCC-ee's projected 0.1\% level measurement of the $ZZh$ coupling would probe $f \sim 6~{\rm TeV}$ and tuning at the part-in-a-thousand level. This scenario is also constrained by $Z$ pole observables, since we expect an $S$-parameter of parametric size $Nv^2/f^2$, but it is a case where the Higgs factory mode shines and sets a more stringent bound.

\begin{figure}
\begin{center}
\includegraphics[width=0.45\textwidth]{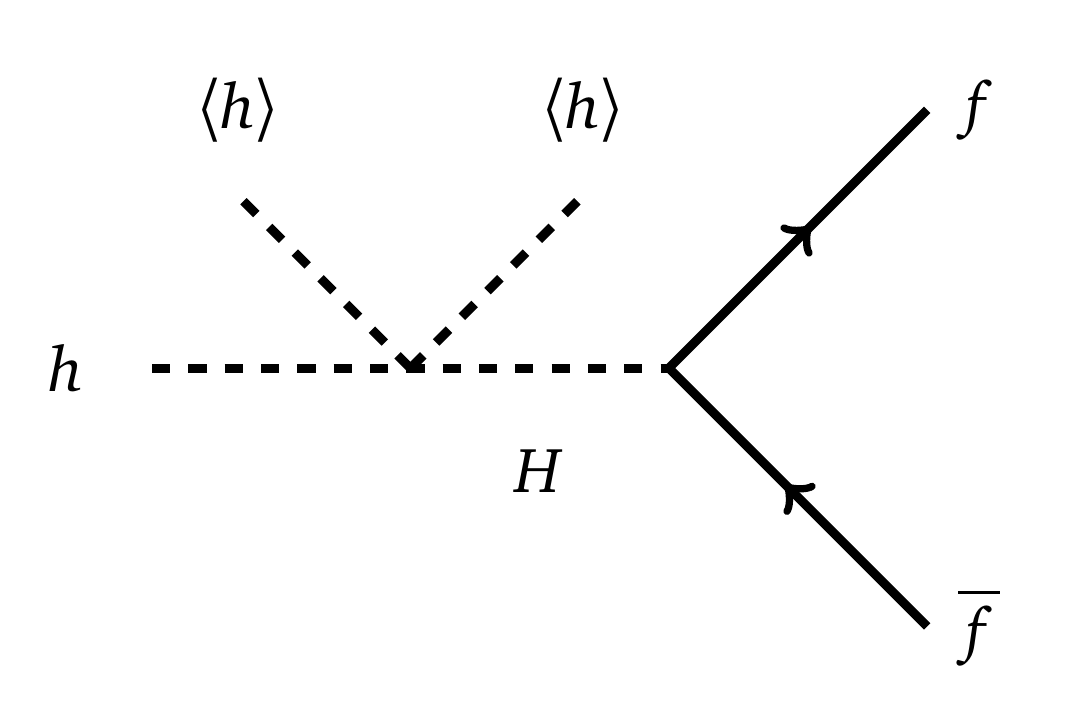}\quad\includegraphics[width=0.35\textwidth]{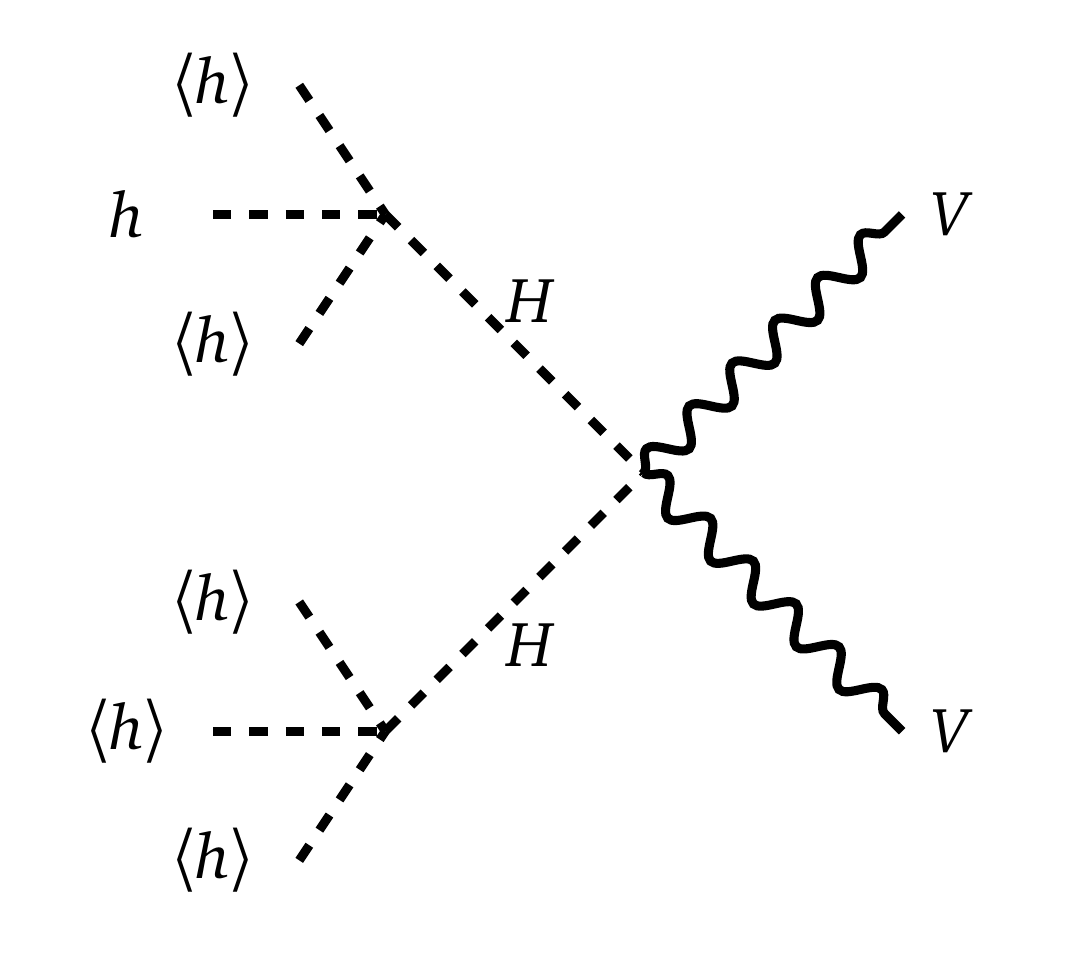}
\end{center}
\caption{Tree-level sources of deviations in couplings of the light Higgs boson in a Two-Higgs Doublet Model. Fermion couplings have larger deviations $\sim m_Z^2/m_H^2$ whereas gauge boson couplings deviate only at order $m_Z^4/m_H^4$.}
\label{fig:2hdmcouplings}
\end{figure}

A very different scenario with tree-level modifications of the Higgs properties occurs in a type-II two-Higgs doublet model (2HDM). A useful way to see the leading deviations in this model is to work in a basis with one doublet $h$ that has a VEV and a second doublet $H$ with no VEV.\cite{Gupta:2012fy} These are not mass eigenstates; rather, the size of the $hH$ and $h^3 H$ terms are linked to ensure that there is no $H$ tadpole. Because the light Higgs we have measured is approximately Standard Model-like, the misalignment between the mass and VEV bases must be small, and we can assess the modifications of the light Higgs couplings by drawing diagrams that mix $h$ and $H$ through VEV insertions, as shown in Figure \ref{fig:2hdmcouplings}. We see that we pay a price of a propagator $1/m_H^2$ in the fermion couplings, but we pay two such factors in the couplings to gauge bosons.\cite{Gunion:2002zf} As a result, a 2HDM model is not probed very effectively through Higgsstrahlung measurements. Rather, the main effect is to modify the fermion couplings, and in particular the largest partial width $\Gamma(h \to b{\bar b})$, through which all other Higgs branching ratios change as well. The roughly percent-level precision achievable on the $hb{\overline b}$ coupling translates into a probe of heavy Higgs boson masses up to beyond 1 TeV, comparable to the long-term reach at the LHC\cite{Djouadi:2015jea,Craig:2015jba,Hajer:2015gka}.

Finally, we mention the case of stops, which when integrated out at one loop generate the Higgs couplings to photons and gluons in equation (\ref{eq:higgsdecay}). The gluon coupling is the most sensitive measurement, which for percent-level accuracy and equal left- and right-handed stop masses translates into sensitivity to masses of about 1 TeV, as depicted in the purple curve of Figure \ref{fig:susycepc}. This is comparable to the electroweak precision reach via the $T$-parameter, although the latter probes only the left-handed stops.

\subsection{Difficult Higgs Couplings}

All of the $e^+ e^-$ colliders can measure the basic Higgs couplings to the $W, Z, \gamma, b, c, g, \tau,$ and $\mu$, as well as the Higgs invisible width. But three other Higgs couplings are of interest and present much greater challenges: those to electrons, to top quarks, and to itself. The Higgs {\em branching ratio} to electrons is much too small to measure. On the other hand, the $s$-channel production of the Higgs, $e^+ e^- \to h$, has a very small but not completely negligible rate\cite{Jadach:2015cwa}. This has led to interesting suggestions to build up very large luminosities (90 ab$^{-1}$) on the Higgs pole at FCC-ee to directly detect the coupling to electrons\cite{d'Enterria:2016yqx}. (I thank Oliver Fischer for bringing up this possibility at the conference.)

The Higgs self-coupling is one of the few remaining quantities in the Standard Model that we have no direct experimental information about. It is absolutely crucial that we measure it accurately in the future in order to learn about the shape of the Higgs potential and possible dynamics underlying electroweak symmetry breaking. The circular $e^+ e^-$ colliders will not be able to directly say anything, though interestingly they are sensitive to order-one deviations via their indirect loop effect on the Higgs cross section \cite{McCullough:2013rea}. If we want truly precise measurements, however, they will come from the circular colliders only when they operate as very high energy hadron colliders. The ILC, on the other hand, can measure the Higgs self-coupling through processes like $e^+ e^- \to Z h h$ at $\sqrt{s} = 500$ GeV and $e^+ e^- \to \nu {\overline \nu} hh$ at $\sqrt{s} = 1$ TeV, where a precision of 16\% may be attained.\cite{Behnke:2013lya,Asner:2013psa}\footnote{For a recent update see the talk of Masakazu Kurata at http://agenda.linearcollider.org/ event/6662/session/30/; I thank Marcel Stanitzki for pointing me to this information.} This again demonstrates that one of the key advantages of a linear collider over a circular collider comes from the higher energies and correspondingly wider range of physics processes that it can access. Along similar lines, we should note that the ILC would be able to directly measure the Higgs boson coupling to top quarks by operating at $\sqrt{s} \approx 500~{\rm GeV}$ (ideally slightly higher in energy) where the process $e^+ e^- \to t{\overline t}h$ can be studied. This could give a $\sim 10$\% direct measurement of the top Yukawa coupling \cite{Yonamine:2011jg,Tabassam:2012it,Asner:2013psa}, which circular colliders can only indirectly constrain through its effect on the Higgs coupling to gluons.

\subsection{Exotic Higgs Decays}

Most of the discussion of physics at $e^+ e^-$ colliders centers around precision Standard Model measurements, which can indirectly probe for new physics. The ILC, by going to a center-of-mass energy of 1 TeV, has more direct discovery potential. But at a circular machine like FCC-ee or CEPC, opportunities for directly finding new particles are limited. A very exciting exception is the possibility of new light particles discovered in rare decays of the Higgs boson. Because the Higgs coupling to $b$-quarks is so small, there is ample room for small couplings to new physics to lead to detectable decay rates. A detailed recent survey of exotic Higgs decays discusses many different channels.\cite{Curtin:2013fra} Possibilities include light pseudoscalars, dark photons, dark matter, and Hidden Valleys.\cite{Strassler:2006im}

To date there has been relatively little study of the possibilities for direct discoveries of exotic Higgs decays at future $e^+ e^-$ colliders. This will be a fruitful area to explore further. For now I just want to make one comment: these decays could involve exotic signals like particles that propagate macroscopic distances in the detector before decaying. It is important when designing detectors that opportunities to see exotic physics aren't unnecessarily closed off---we want to optimize not only for Standard Model measurements, but for discovery potential. Because particles of any given lifetime have a probability distribution of decay locations, it is likely that detectors that can pick up displaced decays in the tracking volume or decays in an outer muon detector, taken together, can cover much of the parameter space, and that discoveries can be made regardless of the zero order choices of detector technology (e.g.~silicon trackers versus TPCs). Nonetheless, the possibilities for exotic signals should be kept in mind throughout the planning process.

\section{Conclusions}

As we have seen above, linear and circular colliders are to some extent complementary. If possible, it would be wonderful to have both. A linear collider has as a major advantage the capability to go to higher energies. It can improve electroweak fits by operating at the $t{\bar t}$ threshold and performing a precise top mass measurement. It could even operate at high enough energy to directly measure the top Yukawa coupling to the Higgs boson. This has great physical significance, as it is the leading interaction driving concerns about fine-tuning in the Standard Model. The linear colliders can also begin to probe the shape of the Higgs potential through self-coupling measurements. Finally, the extra discovery potential for electroweak particles is a major selling point: the ILC operating at $\sqrt{s} = 1$ TeV could discover particles with masses up to about 500 GeV, which in some cases goes well beyond the LHC's expected reach. Circular colliders have a smaller energy reach but an important advantage in energy calibration through resonant depolarization, which can be used to significantly improve our knowledge of the $Z$ boson mass and width. The large luminosities that are proposed at circular colliders also allow extremely precise and powerful tests of the Higgs boson coupling to the $Z$. The most important argument in favor of circular machines is their capacity to be reused as high-energy hadron colliders with immense discovery potential. Given current uncertainties in the magnet technology that we will have in the future, it is important to build any future electron-positron collider with a large enough ring that we can be confident of its future as a hadron collider.

The LHC is a wonderful machine that will tell us a great deal about the existence or nonexistence of new colored particles at the TeV scale, but its abilities to probe particles with only electroweak interactions or even possibly colored particles that decay in ways that mimic backgrounds are limited. Higgs factories will exhaustively probe any new particles that interact with the Higgs, filling gaps that the LHC will leave behind. Electroweak precision tests are very complementary to this. For instance, the $T$-parameter could be the strongest constraint on the model of folded stops, providing one motivation for running at energies below $\sqrt{s} = 240$ GeV for higher accuracy.

My goal in this talk is not to advocate for any particular new collider or run plan, but to emphasize that different physics goals can be optimized by running at different energies or with different types of colliders. We must take all the options seriously. This is an exciting time for particle physics, as we lay the groundwork for future discoveries.

\subsection*{Acknowledgments}
This work was supported in part by the NSF Grant PHY-1415548. It is partly based on work done in collaboration with JiJi Fan and Lian-Tao Wang and on portions of the CEPC pre-CDR to which Zhijun Liang and Jens Erler also made crucial contributions. I would like to thank the organizers at the Hong Kong Jockey Club Institute for Advanced Study for the invitation to speak, and the participants of the meeting for a productive and stimulating environment. I would also like to thank the organizers and participants of the earlier Workshop on Physics at the CEPC in Beijing in August 2015. I would particularly like to thank Alain Blondel, Oliver Fischer, Mike Koratzinos, Maxim Perelstein, Marcel Stanitzki, and Charlie Young for thought-provoking comments and questions at these meetings.

\bibliographystyle{utphys}
\bibliography{higgsfactory}


\end{document}